\documentclass{article}

\usepackage[preprint]{neuralips_custom}

\usepackage[moderate,margins,mathdisplays,mathspacing,lists,indent, floats, paragraphs]{savetrees}
\usepackage{amsmath}
\usepackage{amsthm}
\usepackage{accents}
\usepackage[utf8]{inputenc} 
\usepackage{hyperref}       
\usepackage{url}            
\usepackage{booktabs}       
\usepackage{amsfonts}       
\usepackage{nicefrac}       
\usepackage{microtype}      
\usepackage{xcolor}         
\usepackage{bm} 
\usepackage{xfrac} 
\usepackage{booktabs} 
\usepackage{algorithm}
\usepackage[noend]{algpseudocode}
\usepackage{caption}
\usepackage[margin=20pt]{subcaption}
\usepackage{enumerate}
\usepackage{balance} 
\usepackage{paralist}
\usepackage{dsfont}
\usepackage{rotating}
\usepackage{listings}
\usepackage{adjustbox}
\usepackage{mymacro}
\usepackage{comment}

\usepackage{graphicx}

\title{From Augmentation to Decomposition: A New Look at CUPED in 2023}

\author{%
    \textbf{Alex Deng}$^{1}$ \quad \textbf{Luke Hagar}$^{2}$ \quad \textbf{Nathaniel Stevens }$^{2}$ \quad \textbf{Tatiana Xifara}$^{1}$ \\ 
    \textbf{Lo-Hua Yuan}$^{1}$ \quad \textbf{Amit Gandhi}$^{1}$ \\
    $^1$ Airbnb \quad $^{2}$ University of Waterloo \\
    \texttt{\{alex.deng, tatiana.xifara, lohua.yuan, amit.gandhi\}@airbnb.com}\\
    \texttt{\{nstevens, lmhagar\}@uwaterloo.ca}
}

\begin{document}

\maketitle

\begin{abstract}
Ten years ago, CUPED (Controlled Experiments Utilizing Pre-Experiment Data) \citep{deng2013cuped} mainstreamed the idea of variance reduction leveraging pre-experiment covariates. Since its introduction, it has been implemented, extended, and modernized by major online experimentation platforms \citep{xie2016improving,guo2021machine,poyarkov2016boosted,jin2023toward,expvrblog}. Many researchers and practitioners often interpret CUPED as a regression adjustment \citep{lin2013agnostic,tsiatiscovariateadj}. In this article, we clarify its similarities and differences to regression adjustment and present CUPED as a more general augmentation framework which is closer to the spirit of the 2013 paper. We show that the augmentation view naturally leads to cleaner developments of variance reduction beyond simple average metrics, including ratio metrics and percentile metrics. Moreover, the augmentation view can go beyond using pre-experiment data and leverage in-experiment data, leading to significantly larger variance reduction. We further introduce metric decomposition using approximate null augmentation (ANA) as a mental model for in-experiment variance reduction. We study it under both a Bayesian framework and a frequentist optimal proxy metric framework. Metric decomposition arises naturally in conversion funnels, so this work has broad applicability.  

\end{abstract}


\section{Introduction}
The CUPED method proposed by \citet{deng2013cuped} was inspired by the method of control variates from stochastic simulation \citep{stosim,owen2013monte}. CUPED is a model-free method that relies only on the observation that any pre-experiment difference between two randomized groups is pure noise due to randomization and should be $0$ in expectation. The model-free aspect is at the heart of CUPED; there is no assumption of a relationship of any form between the covariates and the target metric, as long as they have a nonzero correlation to exploit. Also, the authors developed the theory directly on estimators (of metrics and $\Delta$'s), instead of modeling individual subject-level data points like regression models do. However, many blogs and papers citing CUPED often interpret it as a regression adjustment method, focusing on defining it in terms of averages of individual residuals. 

2023 marks a decade since CUPED's initial publication. In this work, we present CUPED as an augmentation framework which is closer to the spirit of the in initial proposal in the 2013 paper. We show that the augmentation view naturally leads to variance reduction beyond simple average metrics; it easily handles ratio metrics and percentile metrics as well. Moreover, the augmentation view can go beyond using pre-experiment data and leverage in-experiment data, leading to significantly larger variance reduction. We further introduce metric decomposition as a mental model for in-experiment variance reduction and present a Bayesian variance reduction framework. Metric decomposition arises naturally in conversion funnels, so this work has broad applicability.  

\paragraph{Notation} Henceforth, we denote the observed metric value from the treatment and control groups as $M_t$ and $M_c$, the unknown true average treatment effect (ATE) as $\delta$, and we use $\widehat{\delta}=M_t-M_c$ to denote the naive estimate of $\delta$ based on the difference between two metric values. This is also denoted as $\Delta$.    

\section{CUPED as Augmentation}\label{sec-aug}

\subsection{Three Simple Insights that Underlie CUPED}
\paragraph{First insight: Augmentation with Mean-Zero Term} For any estimator $\widehat{\delta}$ of $\delta$, define a new estimator
 \begin{equation}\label{eq-vraug1}
 \widehat{\delta}^* = \widehat{\delta} + \widehat{\delta}_0\ 
 \end{equation}
such that $\mathbb{E}\left[\widehat{\delta}_0\right] = 0$ . Then $\widehat{\delta}^*$ is also an estimator of $\delta$ and is unbiased if $\widehat{\delta}$ is. Thus, the CUPED estimator is a mean-preserving augmentation of any existing estimator. The purpose of the augmentation is to find an augmentation such that $
\var\left[\widehat{\delta}^*\right] < \var\left[\widehat{\delta}\right]\ .
$

\paragraph{Second Insight: Optimal Variance Reduction from a Linear Family} 
The second insight of CUPED is that variance reduction is almost guaranteed with any mean-zero augmentation $\widehat{\delta_0}$! This is because any mean-zero augmentation can be multiplied by a scalar $\theta$ yielding a whole family of mean-zero augmentations $\widehat{\delta}^*(\theta) = \widehat{\delta} + \theta\widehat{\delta_0}$, with variance
$
\var\left[\widehat{\delta}^*(\theta)\right] = \var\left[\widehat{\delta}\right]+\theta^2 \var\left[\widehat{\delta}_0\right] + 2\theta \text{Cov}\left[\widehat{\delta}, \widehat{\delta}_0\right] 
$.
This variance is minimized when
\[
\theta \equiv - \frac{\text{Cov}\left[\widehat{\delta}_0, \widehat{\delta}\right]}{\var\left[\widehat{\delta}_0\right]} 
\]
with minimum variance
\[
\var\left[\widehat{\delta}^*(\theta)\right] = \var\left[\widehat{\delta}\right] \times \left(1 - \frac{\text{Cov}\left[\widehat{\delta}_0, \widehat{\delta}\right]^2}{\var\left[\widehat{\delta}\right] \var\left[\widehat{\delta}_0\right]}\right ) 
= \var\left[\widehat{\delta}\right] \times \left(1-\text{Corr}\left[\widehat{\delta}, \widehat{\delta_0}\right]^2\right)\ .
\]
The amount of variance reduction from augmentation by $\widehat{\delta_0}$ is equal to the correlation between the augmentation term and the existing estimator $\text{Corr}\left[\widehat{\delta}, \widehat{\delta_0}\right]^2$. Any mean-zero augmentation is like a direction of gradient descent, and variance is optimally reduced with an appropriate choice of step size. 

\paragraph{Third Insight: Existence of Mean-Zero Augmentation}
The final insight of CUPED is that mean-zero augmentations are abundant if we tap into pre-experiment period data. Let $M(\bX_{pre})$ be any metric computed from a (multivariate) signal $\bX$ on pre-experiment period data (e.g., $\overline{f(\bX)}$ or a percentile, etc.). Then $\Delta_{pre}(M) = M(\bX_{pre})_t - M(\bX_{pre})_c$ is a mean-zero augmentation. 

\subsection{Relation to Regression Adjustment}
When the metric of interest is a simple average and the naive estimator of ATE is the difference in means $\widehat{\delta} = \Delta(Y)$, and the augmentation is also a difference in means $\Delta(Y)^* = \Delta(Y) - \theta \Delta(X)$, CUPED is often compared to regression adjustment of the form
\begin{equation}\label{eq:ancova}
Y_i = \alpha + \delta A_i + \beta X_i  + \epsilon_i \,. \tag{ANCOVA1}
\end{equation}  and 
\begin{equation}\label{eq:ancova2}
Y_i = \alpha + \delta A_i + \beta X_i  + (\gamma X_i)A_i + \epsilon_i \, , \tag{ANCOVA2}
\end{equation} 
where $A_i$ is the treatment assignment assignment indicator. \cite{tsiatiscovariateadj} showed both ANCOVA1 and ANCOVA2 estimate $\delta$ asymptotically by 
$$
\overline{Y_{t}} - \overline{Y_{c}} - (\overline{f(X_{t})}-\overline{f(X_{c})})\, ,
$$
for some function $f$. Therefore, we can see ANCOVA1 and ANCOVA2 are both asymptotically special cases of CUPED. The difference between ANCOVA1 and ANCOVA2 relates to the choice of how to fit the function $f(X)$ using treatment and control data. The linear regression coefficient estimator is $\mathrm{Cov}(X,Y)/\mathrm{Var}[X]$, where the denominator $\mathrm{Var}[X]$ is the same for treatment and control. However, $\mathrm{Cov}[X,Y]$ differs from treatment to control  due to the treatment effect. ANCOVA1 pools the data together and fits a linear regression. In this case we have $$\mathrm{Cov}[X,Y] = p\mathrm{Cov}_T[X,Y] + (1-p)\mathrm{Cov}_C[X,Y]\,,
$$ where $\mathrm{Cov}_T(X,Y)$ and $\mathrm{Cov}_C(X,Y)$ are the covariances in the treatment and control groups respectively. Note that $p$ is the proportion of all units in the treatment group. On the other hand, ANCOVA2 uses $$
\mathrm{Cov}[X,Y] = (1-p)\mathrm{Cov}_T[X,Y] + p\mathrm{Cov}_C[X,Y]\,.
$$

For CUPED, the optimal $\theta$ is 
\begin{equation}\label{eq:opttheta}
\theta^* =  \frac{Cov[\overline{Y_{t}},\overline{X_{t}}]+Cov[\overline{Y_{c}},\overline{X_{c}}]}{Var[\overline{X_{t}}]+Var[\overline{X_{c}}]}\ .
\end{equation}
This $\theta$ asymptotically converges to $(1-p)\mathrm{Cov}_T[X,Y] + p\mathrm{Cov}_C[X,Y]$. Hence CUPED with this arrangement of $\theta$ is asymptotically equivalent to ANCOVA2. Because $\mathrm{Cov}[\overline{Y_{t}},\overline{X_{t}}] = \mathrm{Cov}_T[X,Y]/n_t$ and $\mathrm{Cov}[\overline{Y_{c}},\overline{X_{c}}] = \mathrm{Cov}_C[X,Y]/n_c$, we can see covariances are weighted inversely proportional to the sample sizes. This explains why it is better to weight $\mathrm{Cov}_T[X,Y]$ by $1-p$ and $\mathrm{Cov}_C[X,Y]$ by $p$, instead of using a more straightforward choice of $p$ for $\mathrm{Cov}_T[X,Y]$ and $1-p$ for $\mathrm{Cov}_C[X,Y]$ as in ANCOVA1. From here we also see ANCOVA2 is theoretically better than ANCOVA1, as advocated by \citep{lin2013agnostic}. In practice this difference is small unless $p$ is far from 0.5 and $\mathrm{Cov}_T[X,Y]$ is very different from $\mathrm{Cov}_C[X,Y]$. When $p=0.5$, ANCOVA1 and ANCOVA2 are equivalent. 

\section{Advantages of the Augmentation View}
In the previous section we showed that CUPED is asymptotically equivalent to ANCOVA2 when applied to simple average metrics, and CUPED is also asymptotically equivalent ANCOVA1 when the treatment and control groups are the same size. But the advantage of CUPED is its augmentation view can naturally lead to variance reduction beyond simple average metrics. The augmentation term does not even need to be related to a difference of metric values in the treatment and control groups. We summarize the advantages of the CUPED augmentation view as follows. 

\paragraph{Flexible Metric Form} As an augmentation to any estimator of interest, it is clear that the theory of CUPED doesn't depend on the metric being an average; CUPED can also be applied to percentile metrics and ratio metrics straightforwardly. These are common challenges when practitioners try to implement CUPED with the regression residual interpretation.   

\paragraph{Flexible Augmentation Form} The augmentation $\widehat{\delta_0}$ does not have to be in the form of a difference of two metric values. One recent development of this idea is illustrated in \citet{deng2023zero}, where
the augmentation term $\widehat{\delta_0}$ is constructed from matching and balancing methods in observational causal inference.

\section{Metric Decomposition with Approximately Null Augmentation (ANA) }
We can extend the augmentation view further and consider an augmentation that is not guaranteed to have mean zero. This is motivated by the idea of metric decomposition, where we decompose a metric into two components where treatment effects are believed to be mostly captured in one of the two components. Specifically,  let 
$$M = M_1 + M_2 ~~~ \text{and} ~~~ \Delta(M) = \Delta(M_1)+\Delta(M_2)$$ where the true effect also has the decomposition $\delta = \delta_1+\delta_2$. If $\delta_1$ is close to 0 in most cases, and $M_1$ accounts for a significant portion of variation in $M$, treating $\Delta(M_1)$ as an augmentation and using $\Delta(M_2) = \Delta(M) - \Delta(M_1)$ can yield significant variance reduction. 

Compared to CUPED, we no longer have the guarantee that the augmentation has mean zero. We call this scenario approximate null augmentation (\textbf{ANA}). Let $\vDelta =(\Delta_1, \Delta_2)$ be the decomposed vector of $\Delta(M_1)$ and $\Delta(M_2)$, and let $\vdelta = (\delta_1, \delta_2)$ be the vector of true effect for the two components. Then
$$
\vDelta \sim \vdelta + \veps 
$$
where $\veps$ has known covariance matrix $\Sigma(n)$ and is assumed to be approximately normally distributed due to the Central Limit Theorem. The effect $\vdelta$ follows a bivariate normal distribution with variance-covariance matrix $\Lambda$. Taking an empirical Bayes approach, assuming $\vdelta$ has mean $(0,0)$, we can estimate $\Lambda$ from a set of historical experiments with many realization of $\vDelta$.  Our inference target is $\delta = \delta_1+\delta_2$. 

\subsection{Bayesian Variance Reduction}
The first research question we studied is how decomposing a metric into two components changes the Bayesian posterior distribution for $\delta$. In the simple normal-normal model, assuming $\vdelta$ has mean $0$ and covariance $\Lambda$, we know 
\begin{equation*}
\e \left[\vdelta | \vDelta \right] = S \vDelta\ ~~~ \text{and} ~~~ \var \left[\vdelta | \vDelta\right] = (I - S) \Lambda 
\end{equation*}
where $S = \Lambda (\Lambda+\Sigma)^{-1}$. 
Since $\delta = (1,1)\cdot \vdelta$, we derive the posterior mean to be
\begin{align}\label{eq:bivmean}
\e \left[\delta | \Delta_1, \Delta_2 \right] = C^{-1} \left[(\Lambda_{11}+\Lambda_{12})(\Lambda_{22}+\Sigma_{22}) - (\Lambda_{12}+\Sigma_{12})(\Lambda_{12}+\Lambda_{22})\right ] \Delta_1 \notag \\ 
+ C^{-1} \left[ (\Lambda_{12}+\Lambda_{22})(\Lambda_{11}+\Sigma_{11}) - (\Lambda_{12}+\Sigma_{12})(\Lambda_{11}+\Lambda_{12}) \right] \Delta_2 \, ,
\end{align}
where $C = \Lambda_{11}\Lambda_{22}+\Lambda_{11}\Sigma_{22}+\Lambda_{22}\Sigma_{11}+\Sigma_{11}\Sigma_{22}-\Lambda_{12}^2-2\Lambda_{12}\Sigma_{12}-\Sigma_{12}^2$. 

Alternatively, without the ANA metric decomposition, 
\begin{equation*}
\e \left[\delta| \Delta \right] = A \Delta\ ,\quad  \var \left[\delta | \Delta\right] = A \sigma^2  \ ,
\end{equation*}
where $A = \frac{\lambda^2}{\lambda^2+\sigma^2}$ with $\lambda^2 = \Lambda_{11}+\Lambda_{12}+\Lambda_{21}+\Lambda_{22}$ and $\sigma^2 = \Sigma_{11}+\Sigma_{12}+\Sigma_{21}+\Sigma_{22}$. ANA metric decomposition naturally leads to variance reduction under the Bayesian framework. We proved that 
$$
1^T\cdot (I - S) \Lambda \cdot 1 < A \sigma^2\, .
$$
The proof is omitted here but we will demonstrate the finding using empirical results.

\subsection{Frequentist Optimal Proxy Metric with Variance Reduction}
As an extension of CUPED, ANA can be used as a frequentest estimator and analyzed as a proxy metric of the form $\Delta^* := \Delta_2 + \theta \Delta_1$. Comparing to the Bayesian posterior mean, the main difference is that we do not put a shrinkage factor on the signal component $\Delta_2$, and only shrink the ANA component $\Delta_1$. 

\begin{thm}
Among ANA estimators $\Delta^* := \Delta_2 + \theta \Delta_1$, the mean squared error $\e [(\delta - \Delta^*)^2]$ is minimized when
\begin{equation}\label{eq:minerr}
\theta = \frac{\Lambda_{11}-\Sigma_{12}}{\Lambda_{11}+\Sigma_{11}}\, .
\end{equation}
The correlation between $\delta$ and $\Delta^*$ is maximized when
\begin{equation}\label{eq:maxcorr}
\theta = \frac{(\Lambda_{12}+\Sigma_{12})(\Lambda_{12}+\Lambda_{22})-(\Lambda_{11}+\Lambda_{12})(\Lambda_{22}+\Sigma_{22})}{(\Lambda_{12}+\Sigma_{12})(\Lambda_{11}+\Lambda_{12})-(\Lambda_{12}+\Lambda_{22})(\Lambda_{11}+\Sigma_{11})} \, .
\end{equation}
\end{thm}

\paragraph{Minimizing Effect Prediction Error} The ANA estimator with \eqref{eq:minerr} minimizes the effect prediction error. It is a generalization of CUPED in the sense that when $\Lambda_{11}=0$, $\theta = -\Sigma_{12}/\Sigma_{11}$ reduces to CUPED.

\paragraph{Maximizing Correlation} \citet{tripuraneni2023choosing} propose an alternative objective in which interest lies in maximizing the correlation between the true effect and the estimate. Comparing \eqref{eq:maxcorr} to \eqref{eq:bivmean}, we see that the ANA estimator $\Delta_2 + \theta \Delta_1$ that maximizes the correlation between $\delta$ and $\Delta^*$ is simply a rescaled Bayesian posterior mean estimator such that $\Delta_2$ receives no shrinkage. 

\section{Empirical Results}\label{sec:empirical}
We applied Approximate Null Augmentation to 25 early stage ranking experiments at Airbnb. These early stage experiments run for roughly 1 week taking a small percentage of total traffic. The main target metric of interest is booking per guest. To construct the ANA, we leverage counterfactual ranking results. That is, for each search, we compare the ranked results produced by the treatment and control ranker. If a click on a booked listing is ranked in close proximity according to both rankers, or if a click is from the map and both rankers would show the listing on the map, then the attributed booking from this click would be approximately the same regardless of the treatment assignment. In addition, we use a utility model to attribute a user's booking to searches \citep{deng2023variance}. The end result is for each booking, we can construct an ANA component representing a fraction of the booking that both rankers would have contributed almost equally. 

The effect covariance $\Lambda$ and the average covariance of the noise $\veps$ were estimated to be (after scaling by the same constant)
$$
\Lambda = 
\begin{pmatrix}
0.576 & -0.896\\
-0.896 & 4.329 
\end{pmatrix} \quad \text{and} \quad
E[\Sigma] = 
\begin{pmatrix}
4.020 & 0.169\\
0.169 & 0.811 
\end{pmatrix}\ .
$$
We find the first ANA component $\Delta_1$ displays a variance 5 times larger than the second component $\Delta_2$; while the variance of the effect for the ANA component is less less than 1/7 of $\Delta_2$. If the ANA component has theoretical mean $0$, then the variance of the effect should also be $0$. A close to $0$ effect variance and a large noise variance (both relative, comparing to $\Delta_2$) mean CUPED using ANA can lead to significant variance reduction with a small bias trade-off. 

\begin{table}[!hb]
\centering
\begin{tabular}{@{}lllll@{}}
\toprule
$\Delta$ & $\Delta_1$ & $\Delta_2$ & $\text{ANA}_{corr}$ & $\text{ANA}_{err}$ \\ \midrule
4/25     & 2/25       & 8/25       & 8/25           & 8/25          \\ \bottomrule
\end{tabular}
\caption{Number of statistically significant results out of 25 experiments. CUPED with ANA maximizing correlation and minimizing error has the same number of significant results as using Component 2 only ($\theta=0$).}
\label{tab:stat_sig_tab}
\end{table}

\begin{figure}[thp!]
    \centering
    \includegraphics[width=0.7\textwidth]{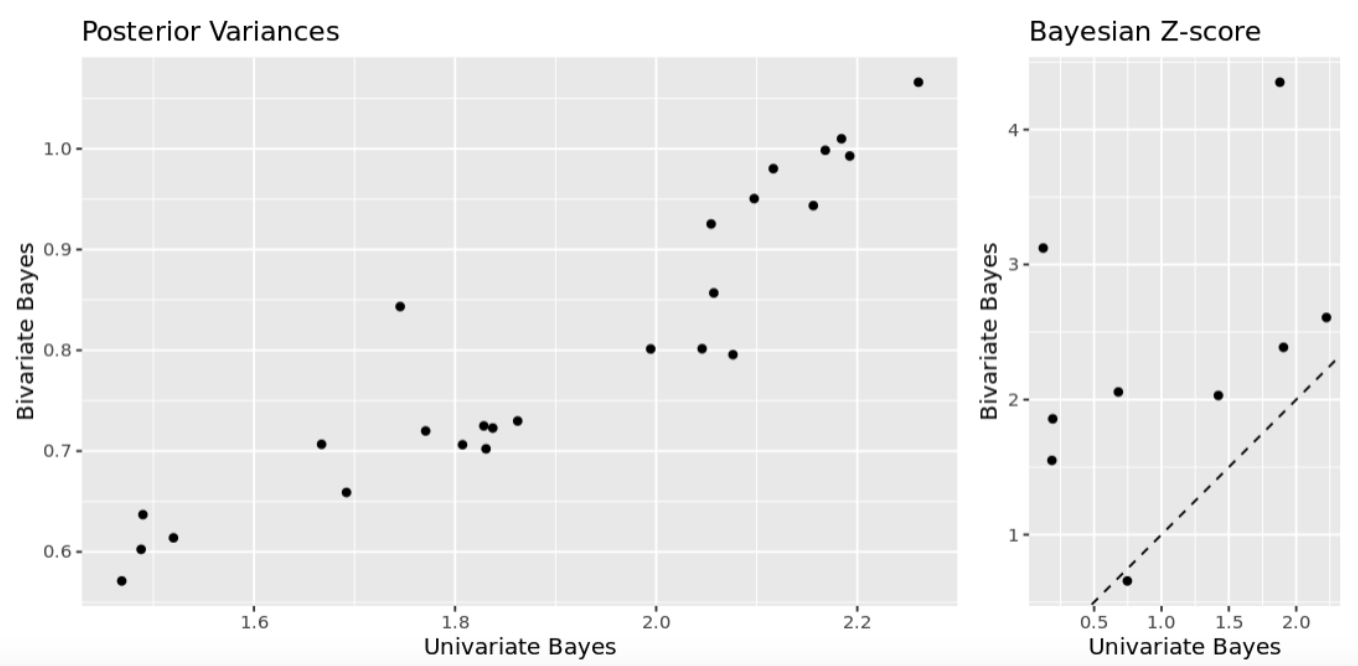}
    \caption{Left: Bayesian posterior using bivariate decomposed model is smaller than without decomposition. Right: Bivariate model produces larger absolute Bayesian Z-score. }
    \label{fig:bayes}
\end{figure}

\begin{figure}[thp!]
    \centering
    \includegraphics[width=0.6\textwidth]{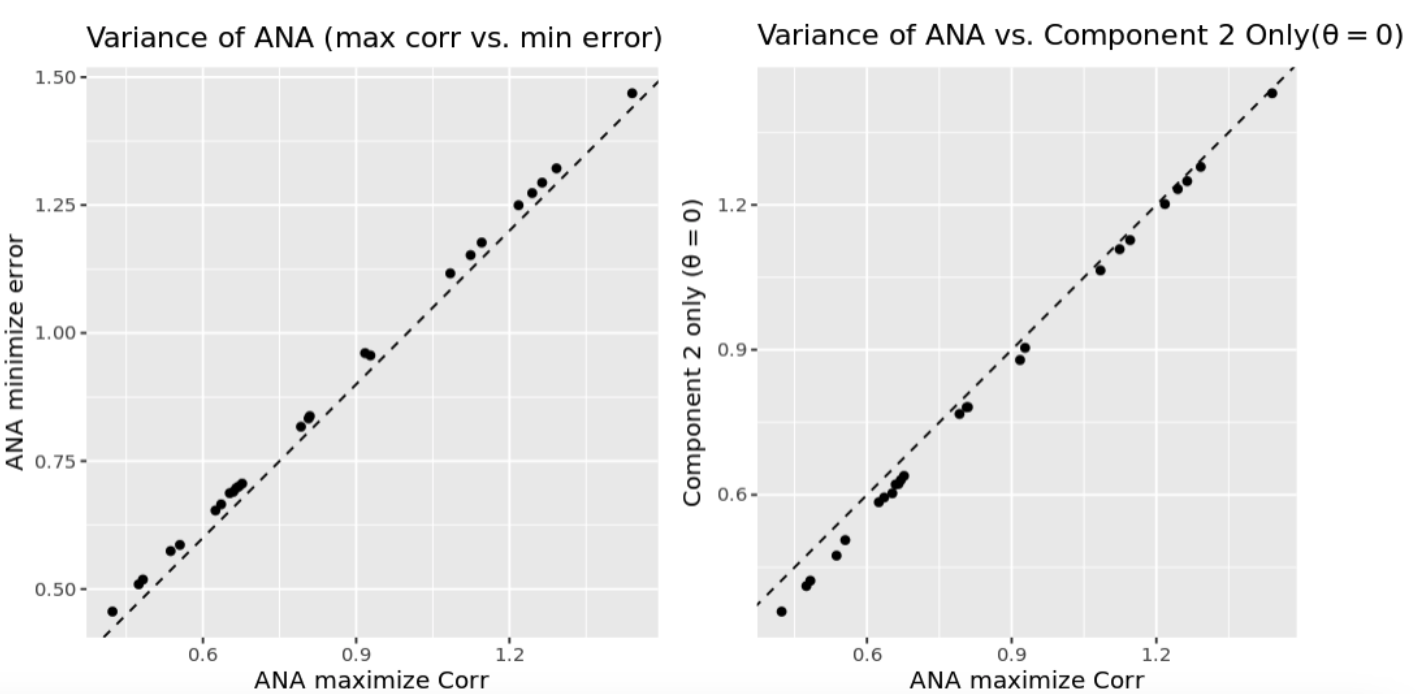}
    \caption{ANA maximizing correlation has a slightly smaller variance than ANA minimizing error (Left); but slightly larger variance than $\Delta_2$ (Right). Difference in variances in these 25 experiments are all very small.}
    \label{fig:freq_var}
\end{figure}

Table~\ref{tab:stat_sig_tab} demonstrates that using ANA with $\theta=0$ ($\Delta_2$), or maximizing correlation, or minimizing error all lead to more statistically significant results, compared to the original $\Delta$ before decomposition. Figure~\ref{fig:bayes} shows that the posterior variance is greatly reduced with the bivariate model. Furthermore, the Bayesian Z-score (posterior mean divided by posterior standard deviation) has a greater absolute value under the bivariate model. Figure~\ref{fig:freq_var} compares variances of ANA with $\theta=0$ ($\Delta_2$), ANA maximizing correlation, and ANA minimizing error. All three produces similar variances. 


\bibliographystyle{ACM-Reference-Format}
\bibliography{ref}


\end{document}